\documentclass[twoside,11pt]{article}

\usepackage{jmlr2e}
\usepackage{amsmath}

\begin{document}

\title{Evaluating Gaussian processes for sparse irregular spatio-temporal data}

\author{\name Mehmet S{\"u}zen \thanks{Correspondence E-mail: mehmet.suzen@physics.org} and Abed Ajraou}

\editor{na}

\maketitle

\begin{abstract} 
A practical approach to evaluate performance of a Gaussian process regression models (GPR) 
for irregularly sampled sparse time-series is introduced. The approach entails 
construction of a secondary autoregressive model using the fine scale predictions 
to forecast a future observation used in GPR. We build different GPR models for Ornstein-Uhlenbeck and Fractional processes for simulated toy data with different sparsity levels to assess the 
utility of the approach.  
\end{abstract}


\section{Introduction}

Time-series analysis is very common subject that manifest itself in sciences and in industry 
[\cite{hamilton1994time}]. Temporal data rarely available in regular intervals and with sufficient sample size, i.e., sparse, irregularly occured and noisy [\cite{richards2011a}]. In these circumstances, standard 
modelling techniques would not be appropriate. Gaussian proceses provide a powerful alternative 
[\cite{williams2006a}], where a prior knowledge can be used without any restrictions on regularity or sparsity 
on the temporal data [\cite{roberts2013a}]. But a measure of goodness of fit would be an issue in this setting. A usual approach is to measure goodness of fit by comparing the results based on a gold standard result. Here we propose an approach to measure performance of GP without resorting to a gold standard result in sparse time-series via building a secondary model based on the resulting.
 
\section{Gaussian Processes}

A short sketch of the machinery of the Gaussian Processes [\cite{williams2006a}] is presented as follows. The starting point for Gaussian process is to define an arbitrary function that explains the outcome with a 
variance $\sigma_{y}$ and noise $\epsilon_{t}$,
$$y(t) = f(t) + \sigma_{y} \epsilon_{t}$$

The primary approach in GP is that training time-series $(y,x)$ and $(y^{*},x^{*})$ the time-series that is to be learned can be expressed in a joint distribution, which is a multivariate Gaussian distribution, 

\begin{displaymath}
p(y,y^{*}) = \mathcal{N} \Big(
     \begin{bmatrix}
      x        \\
      x^{*} 
     \end{bmatrix},
     \begin{bmatrix}
       K_{xx}      &  K_{xx^{*}}     \\
       K_{x^{*}x}  &  K_{x^{*}x^{*}}
     \end{bmatrix}
     \Big)
\end{displaymath}
where $x \in \mathbb{R}^{n x b}$ and $x^{*} \in \mathbb{R}^{m x d}$, where $n$ and $m$ are number of rows 
and $d$ is the number of predictors, the Kernel matrices can be computed as follows, in multivariate setting,
\begin{displaymath}
  K_{i,j}(x_{i}, x_{j}) = - \sigma^{2} \exp\Bigg(-\beta \sum_{k=1}^{d} \Big( \frac{x_{i}^{k}-x_{j}^{k}}{l_{d}} \Big)^{\alpha_{d}}\Bigg)
\end{displaymath}
$i=1,..,n$ and $j=1,..,m$, where $d$ is the number of features or variates at each time point. Kernel choice here is not unique but this is a generalised form of the square exponential kernel. Hyperparameters $(\sigma^{2}, \beta, l, \alpha )$ can be interpreted as signal variance, scaling factor, length scale and roughnes on the time-series respectively. Hence, GP as a function approximation can be obtained in a closed form,  the mean function and the covariance matrix,
\begin{align*}
L           & = (K_{xx} + \sigma^{2}\mathbb{I})^{-1} \\
\bar{y^{*}} & = K_{x^{*}x} L^{-1} y \\
y^{*}_{cov} & = K_{x^{*}x^{*}}-K_{x^{*}x} L^{-1} K_{xx^{*}}
\end{align*}

Note that $y^{*}$ is the observations that is to be learned from the training data.

\section{Evaluation Technique}

A variable of interest $y$ appears regularly over time. A temporal evolution of $y$ can be expressed with ordered set $\Omega^{*}=(t^{*}_{i}, y^{*}_{i})$, where $i=1,..,n$. Only a subset of $\Omega^{*}$ may be observed in irregularly spaced intervals and not too frequently, i.e., sparse. These sparse observations of $y$ appear in the subset $\Omega=(t_{j}, y_{k})$,  where $\{k \in \{1,..n\}\}$, as an ordered sequence but irregular, i.e., irregular time-series. Building a Gaussian Process model to construct the original time series $\Omega^{*}$ using the partial information $\Omega$ is one of the most promising approach for sparse irregular temporal data [\cite{richards2011a}]. The resulting series can be denoted by $\Omega^{*}_{gp}=(t^{*}_{i}, Y^{*}_{i})$. 

Performance of the resulting reconstruction $\Omega^{*}_{gp}$ usually measured against a gold standard method [\cite{roberts2013a}]. We propose using a secondary Autoregressive Model (AR) based on the resulting set $\Omega^{*}_{gp}$ is proposed. Construction of $n-1$ different autoregressive models to predict observation points $y_{j}$. Procedure is as follows.

\begin{enumerate}
 \item Fixed a horizon $h$, for Autoregressive model prediction.
 \item Select the $m-1$ subsets of $\Omega^{*}_{gp}$ each up to a point $y_{j}$, $j_{k}-h$, denoting each subset, $(\Omega^{*}_{gp})_{k}$, where $k=2,..,m$. 
 \item Build an AR model for each $(\Omega^{*}_{gp})_{k}$ and predict next $y_{k}$ as $y_{k}^{ar}$.
 \item Mean Absolute Percent Error (MAPE-AR) can be computed, $M = \frac{1}{m-1} \sum^{k=2}_{m} (y_{k}-y_{k}^{ar})/y_{k}$  
\end{enumerate}

MAPE-AR value will quantify the goodness-of-fit for GP regression without resorting to a gold standard method.

\subsection{Simulated Data}

\begin{figure}
  \includegraphics[width=0.5\textwidth]{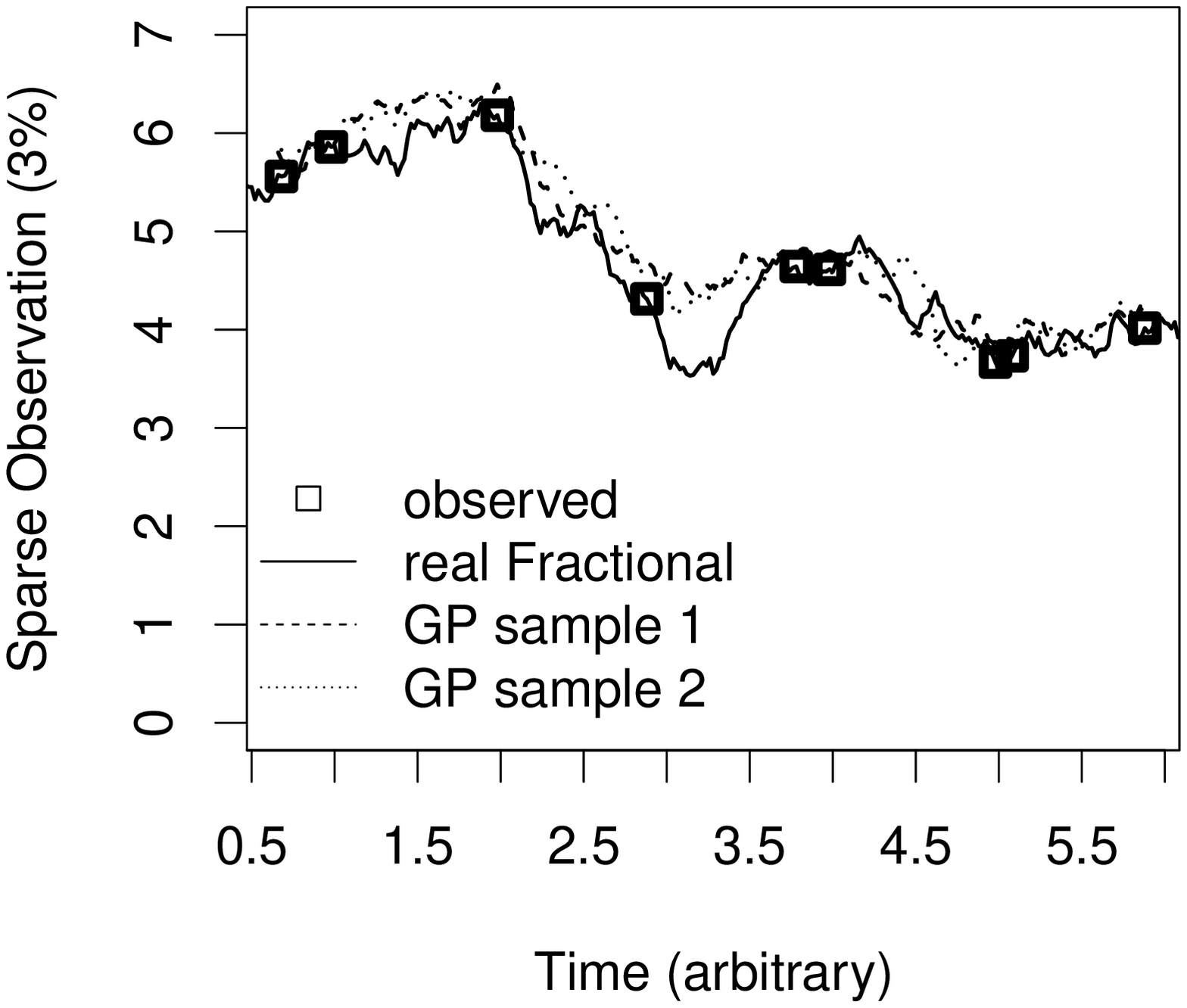}  
  \includegraphics[width=0.5\textwidth]{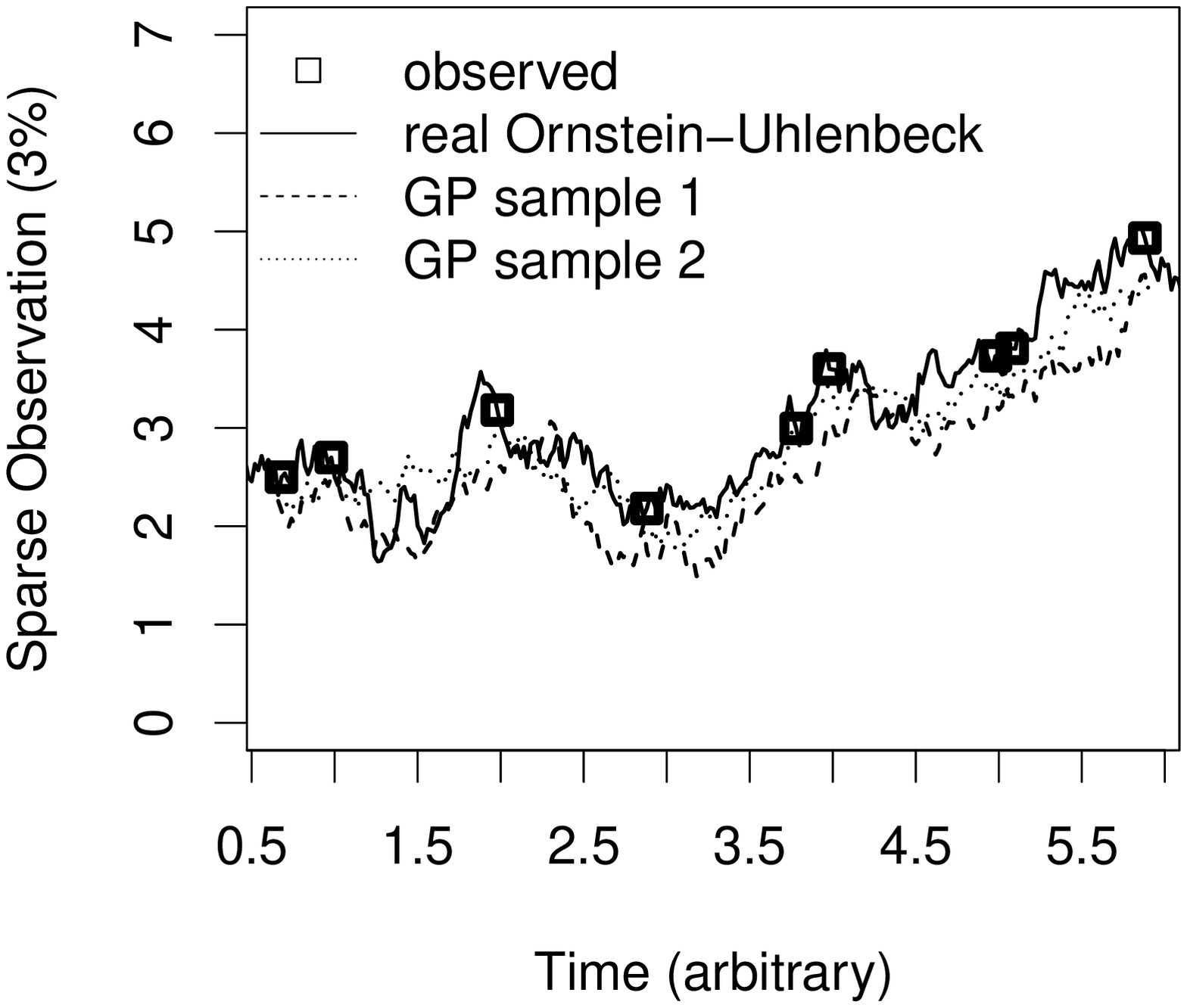}
  \caption{Simulated Ornstein-Uhlenbeck and fractional processes and with $3$ percent sparsity level. Two functions are drawn from the resulting GP regression.}
\end{figure} 

A pair of toy data is generated, using generalised form of the square exponential kernel, Ornstein-Uhlenbeck and Fractional process with Kernel hyperparameters $(\alpha, \beta, l, \sigma)$, $(1.0, 1.0, 2.0, 1.0)$ and $(1.3, 1.0, 2.0, 1.0)$ respectively. Simulated data contains $351$ observations points with regular time spacing of $0.02$. New subset of observations generated by randomly selecting $3$, $5$ and $7$ percent of the simulated data, at least $5$ time-steps apart. This constitutes different sparsity levels.

\subsection{Experiments} 

We fit Gaussian Process on these sparse data sets using square exponential kernel. Results for $3$ percent sparse data sets are shown on the Figure 1.

Using seasonal $ARIMA(1,1,1)(1,1,1)$ as a secondary autoregressive model, MAPE-AR measure is summarized in Figure 2, for demonstration purposes.

\begin{figure}
 \centering
 \begin{tabular}{ | l | l | c | r | }
   \hline
                       & 3\% sparsity   &  5\% sparsity & 7\% sparsity    \\
   \hline
   Ornstein-Uhlenbeck  & 0.0049         & 0.0051        & 0.0067 \\
   Fractional          & 0.0042         & 0.0056        & 0.0057 \\
   \hline
 \end{tabular}
  \caption{MAPE-AR measure of goodness-of-fit results for different sparsity levels.}
\end{figure} 

\section{Summary}

A technique to measure goodness-of-fit in Gaussian processes for sparse temporal data is proposed based on building secondary autoregressive model to construct the regularly space data. In our emprical investigation we have demonstrated the utility of the approach using simulated data. 





 
\bibliography{gp}

\end{document}